\documentclass[nocopyright]{cimento_arxiv}

\usepackage{graphicx}
\usepackage{url}
\usepackage{eurosym}

\graphicspath{{figures/}}

\newcommand{\arduino}{Arduino}
\newcommand{\plasduino}{Plasduino}

\title{\plasduino: an inexpensive, general purpose data acquisition framework
  for educational experiments}

\shorttitle{\plasduino: a data acquisition framework for educational
  experiments}

\author{%
  L.~Baldini\from{ins:unipi}\from{ins:infn}\ETC,
  C.~Sgr\`o\from{ins:infn},
  E.~Andreoni\from{ins:unipi},
  F.~Angelini\from{ins:unipi},
  A.~Bianchi\from{ins:unipi},
  J.~Bregeon\from{ins:cnrs},
  F~.Fidecaro\from{ins:unipi},
  M.~M.~Massai\from{ins:unipi},
  V.~Merlin\from{ins:unipi},
  J.~Nespolo\from{ins:unipi},
  S.~Orselli\from{ins:unipi}
  \atque
  M.~Pesce-Rollins\from{ins:infn}
}
\instlist{%
  \inst{ins:unipi} Dipartimento di Fisica E.~Fermi, Universit\`a di Pisa
  - Pisa, Italy
  \inst{ins:infn} INFN, Sezione di Pisa - Pisa, Italy
  \inst{ins:cnrs} CNRS IN2P3/INSU - Montpellier, France
}

\PACSes{\PACSit{01.40.-d}{Education}
\PACSit{01.50.Pa}{Laboratory experiments and apparatus}
\PACSit{07.05.Hd}{Data acquisition: hardware and software}}

\begin{document}

\maketitle

\begin{abstract}
Based on the \arduino\ development platform, \plasduino\ is an open-source data
acquisition framework specifically designed for educational physics experiments.
The source code, schematics and documentation are in the public domain under a
GPL license and the system, streamlined for low cost and ease of use, can be
replicated on the scale of a typical didactic lab with minimal effort.

We describe the basic architecture of the system and illustrate its potential
with some real-life examples.
\end{abstract}

\section{Introduction}

\plasduino~\cite{ref:plasduino} is a project for a data acquisition framework
aimed at educational physics experiments. In brief, the basic goals driving the
development effort can be summarized in three simple adjectives: \emph{simple},
\emph{inexpensive}, \emph{open}.

In this paper we provide a top-level description of the system, characterize
its basic performance and describe a few illustrative use cases.

\subsection{Simple as: easy to set up, easy to use, easy to extend}

In this context we use the adjective \emph{simple} referring to three somewhat
different aspects of the system, namely (i) its set-up, (ii) its use and (iii)
its modification. Most of the dedicated commercial solutions that can be bought
off the shelf are simple in the sense of (i) and (ii) but cannot be modified.
On the other hand, most of the custom (and sometimes very \emph{creative}, be
this good or bad) solutions frequently encountered in didactic laboratories
are seldom easy to set up and use. To this respect, we aim at getting the best
of both worlds.

\subsection{Inexpensive as: it doesn't get much cheaper than that}

Cost is clearly a crucial aspect in terms of creating a framework that can be
replicated and used on the scale of a didactic lab.
To give some context---assuming that one has access to a personal computer with
a reasonably recent operating system---our project aims at providing all the
components to assemble a general-purpose data acquisition system (including the
control hardware, some basic sensors and a fully fledged suite of pre-compiled
applications) for under $50$~\euro.

\subsection{Open as: free for you to use, study and modify}

The concept of \emph{openness} in the context of computer software typically
ties to the right of the users to run, distribute, study and change the software
itself. As a matter of fact, most of these notions apply to an electronic board
no less than a piece of software. To this respect, we are committed to provide,
with no restriction, all the knowledge base that is needed to study, understand
and modify the system, including all the hardware schematics and the source
code of the software.

To some extent we consider this third design goal as the single most relevant
feature of the project. While we fully realize that only a (small) subset
of the users can possibly have the expertise to fully take advantage of it,
this approach is undoubtedly closer to that of the \emph{professional} 
experimental physicist and opens interesting perspectives for additional and
diverse educational paths.

\section{Overview of the project}

The basic system concept described here consists of a small data acquisition
board able to read signals from various sensors and connected with a standard
personal computer through a USB cable. There are four main components to the
framework, namely:
(i) an \arduino\ board, equipped with a microcontroller;
(ii) the \emph{firmware}, i.e., the computer program running on the 
microcontroller itself;
(iii) a series of custom \emph{shields}, or printed-circuit boards acting
as an interface between the microcontroller and the external sensors;
(iv) the control software running on the host computer.
In this section we shall briefly review these basic components.

\subsection{The \arduino\ platform}

``Arduino is an open-source electronics prototyping platform based on flexible,
easy-to-use hardware and software. It's intended for artists, designers,
hobbyists and anyone interested in creating interactive objects or
environments.''
Although this concise, top-level description~\cite{ref:arduino} by the
\arduino\ developers seems to have a very loose connection with the concept of
a real-time data acquisition system, we shall see in a moment that the platform
has a number of appealing features for our purposes.

The \arduino\ team provides a series of affordable ($20$--$50$~\euro)
boards. Peculiarities aside, all of them feature at least: a microcontroller;
$14+$ digital I/O pins---some or all of which have pulse-width modulation (PWM)
capabilities and/or can handle interrupts; $6+$ I/O analog pins with a $10$ or
$12$~bit ADC; and a serial interface to communicate, in both directions, with a
host computer via a standard USB cable.

Compilers exists that can translate high-level C or C++ code into assembled
binary code to be uploaded on the microcontroller and the \arduino\ core
library implements an abstraction layer acting as a common interface to the
hardware and hiding the (sometimes intricate) internals of the hardware itself.
At the cost of some overhead in terms of performance, this layer makes it
trivial to perform the most common operations (addressing the I/O pins,
latching the internal timer, attaching interrupt service routines to digital
inputs and controlling serial devices). Such operations can be streamlined for
speed, if needed, but we shall see in the following that this is effectively
seldom required for our purposes.

\subsection{The \arduino\ firmware}

Building on top of the \arduino\ core library, we have implemented a set of
configurable programs (or \emph{sketches}) for the microcontroller, e.g., for
time measurement or for sampling signals at the analog inputs.
In addition to the source code of the sketches we provide a set of
corresponding binary files compiled for the most popular \arduino\ boards and
our software infrastructure provides all the functionalities for an automatic
upload on the microcontroller, so that this part of the framework is
effectively transparent to the end users.

\subsection{The custom shields}

The \emph{shields} are small PCB to be plugged on top of the \arduino\ board and
acting as an interface between the board and the external world (e.g., with
sensors) or extending the functionalities of the board itself.

In the context of \plasduino\ we are actively developing a series of custom
shields tailored at different educational experiences.
The reader is referred to the \plasduino\ website~\cite{ref:plasduino} for all
the necessary information to replicate our shields. We provide etchable pdf
files for in-house PCB prototyping, Gerber files for industrial production and
detailed data sheets including the part list and a detailed description of the
basic functionalities. While replicating our shields is the single most
difficult step in getting \plasduino\ up and running, this is hardly a
prohibitive task even for novices.

\subsection{The software infrastructure}

The software infrastructure we have developed to control the data acquisition
is entirely written in the Python~\cite{ref:python} programming language and
relies on a number of free and open source packages---most notably the PyQt
framework~\cite{ref:pyqt} for the graphical user interface. All the system
components are intrinsically cross platform and, while we only provide
automated installation packages for the most popular GNU/Linux distributions,
\plasduino\ has been successfully run under Windows. In addition to that, it
should be relatively straightforward to port it to Mac OS and even to newer
low-cost hardware platforms such as UDOO~\cite{ref:udoo}.

The core of the \plasduino\ software framework is a multi-threaded application
for the control of the data acquisition. Among its salient features are:
auto-detection of the \arduino\ board, automatic upload of the microcontroller
firmware and handling of data collection, processing and archival. The system
provides a flexible graphical user interface~\cite{ref:gui} from which the user
can select the specific application to run, configure the settings, start
and stop the data acquisition. In addition to that we do provide a library
of sensors classes (handling the conversion between ADC counts and physical
units) and a set of device classes implementing the serial communication
protocol for specific integrated circuits.

\section{Basic calibration}

In this section we describe some of the calibration studies we have
performed on the system.

\subsection{Time measurements}

The standard \arduino\ library provides a function to latch the value of a
32~bit internal counter incremented by the system clock prescaled by a factor
$64$. The \arduino\ board we used being clocked at $16$~MHz, this corresponds
to a nominal granularity of the time measurements of $4~\mu$s---and a time
resolution of $4/\sqrt{12} \sim 1.15~\mu$s. Though it is possible, in
principle, to reduce the prescale factor in order to improve the timing
performance, we shall see in a moment that this is seldom the actual limiting
factor in real-life applications.

We characterized the timing capabilities of the system by feeding the 1-Pulse
per second (PPS) signal from a GPS receiver into one of the digital inputs of
an \arduino\ board. We used an interrupt to latch the value of the timer on the
rising edge of the PPS signal. 
In order to cope with the $\sim 10^{-5}~^{\circ}$C$^{-1}$ temperature coefficient
typical of ceramic resonators (such as the one used by the \arduino\ board
under test to generate the $16$~MHz clock signal) we performed the timing
calibration in a climatic chamber and we continuously monitored the temperature
of the resonator itself by means of a thermistor connected to one of the
\arduino\ analog inputs.

\begin{figure}[!htb]
  \begin{minipage}[b]{0.45\textwidth}%
    \caption{Average deviation of the measured time interval between two
      consecutive PPS pulses from a GPS receiver with respect to the nominal
      value of 1~s as a function of the temperature. A linear fit between
      $\sim 25~^\circ$C and $\sim 45~^\circ$C returns a value for the slope of
      $-14.0 \pm 0.3~\mu$s~$^{\circ}$C$^{-1}$.}
    \label{fig:gps_1pps_trend}
  \end{minipage}\hfill%
  \includegraphics[width=0.5\textwidth]{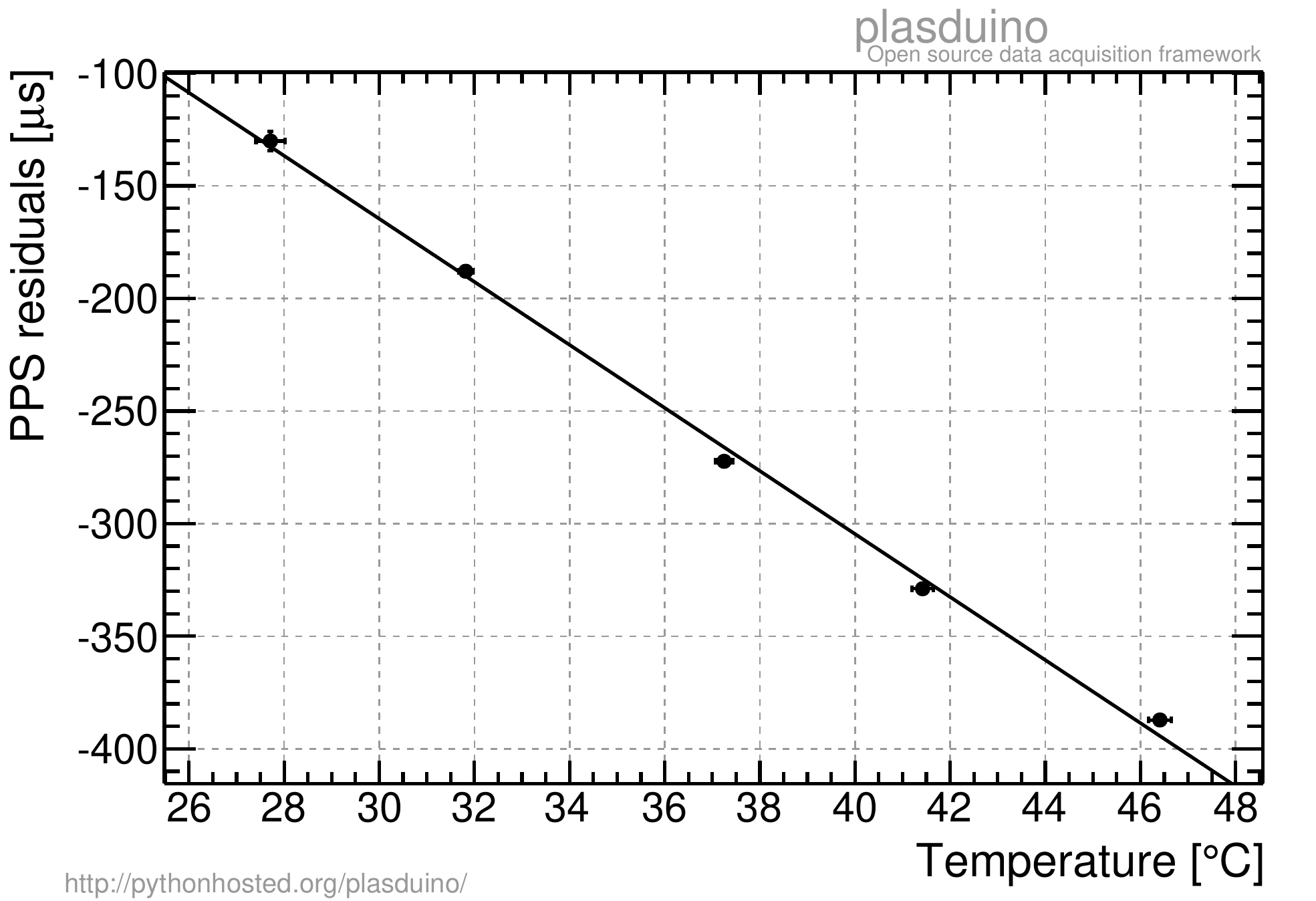}
\end{figure}

Figure~\ref{fig:gps_1pps_trend} shows the average deviation of the measured
time interval between two consecutive PPS pulses with respect to the nominal
value of 1~s as a function of the temperature. The negative slope of
$\sim 14~\mu$s~$^{\circ}$C$^{-1}$, in line with the aforementioned temperature
coefficient, implies that the temperature must be controlled to
$\sim 0.1~^{\circ}$C or better in order to fully exploit the intrinsic time
resolution of the system---unless the temperature itself is monitored and
used for an off-line, board-by-board correction. Integrating a GPS receiver
onto a shield constitutes a viable and effective (though slightly more
expensive and significantly more complex) alternative. We note, however, that
a $\mu$s-type timing resolution is seldom required in typical mechanics
educational experiments.

\begin{figure}[!htb]
  \includegraphics[width=0.5\textwidth]{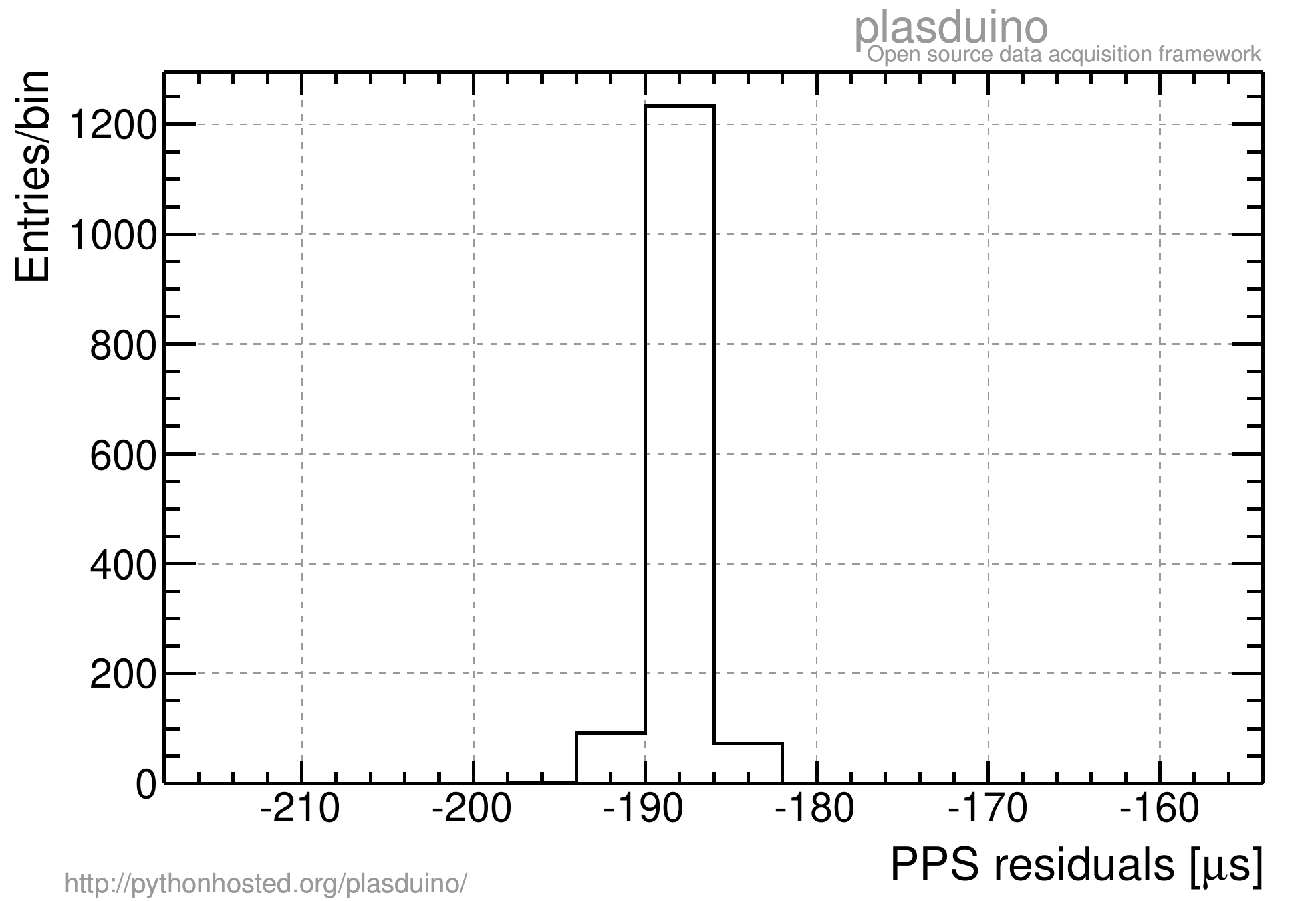}\hfill%
  \begin{minipage}[b]{0.45\textwidth}%
    \caption{Deviations from the nominal value ($1$~s) of the measured time
      intervals between two successive 1-PPS pulses from a GPS receiver.
      The measurement was performed in a climatic chamber with the temperature
      maintained at $31.8 \pm 0.1~^\circ$C. The mean value of the histogram
      corresponds to an absolute time accuracy of $\sim 1.9 \times 10^{-4}$
      and its standard deviation is $1.39~\mu$s, close to the nominal time
      resolution of the system.}
    \label{fig:gps_1pps_dev}
  \end{minipage}
\end{figure}

For completeness figure~\ref{fig:gps_1pps_dev} shows the distribution of the
measured time deviations for a fixed temperature. The root mean square of the
histogram bins ($1.39~\mu$s) is comparable to the nominal time resolution of
the system.

Though we have not performed systematic studies on a large sample of boards,
we regard the figures quoted in this section as representative of the typical
uncertainties. Without any specific calibration effort the system provides
an uncertainty on the absolute time scale of a few parts in $10^4$ and a
time resolution of the order of a few tens of $\mu$s or better (conservatively
assuming that the ambient temperature is controlled to a few $^\circ$C during
the measurements).

\subsection{Temperature measurements}

We briefly discuss temperature measurements as a typical use case for the
\arduino\ analog inputs. In this case the overall performance depends on the
sensor in use; here and in the following, for the sake of discussion, we shall
take as an example the NXFT15XH103FA2B, a $10$~k$\Omega$ negative temperature
coefficient thermistor readily available for (much) less than 1~\euro\ that we
routinely use in our didactic laboratories.

\begin{figure}[!htb]
  \begin{minipage}[b]{0.45\textwidth}%
    \caption{Contributions to the temperature measurement error in our
      reference setup (a NXFT15XH103FA2B thermistor with a $10$~k$\Omega$
      resistor in series). The stochastic term from the ADC resolution
      typically determines the differential accuracy. Being highly positively
      correlated, the other contributions only affect the overall absolute
      temperature scale, i.e., in a limited temperature range they produce
      a rigid vertical shift.}
    \label{fig:tempmon_resolution}
  \end{minipage}\hfill%
  \includegraphics[width=0.5\textwidth]{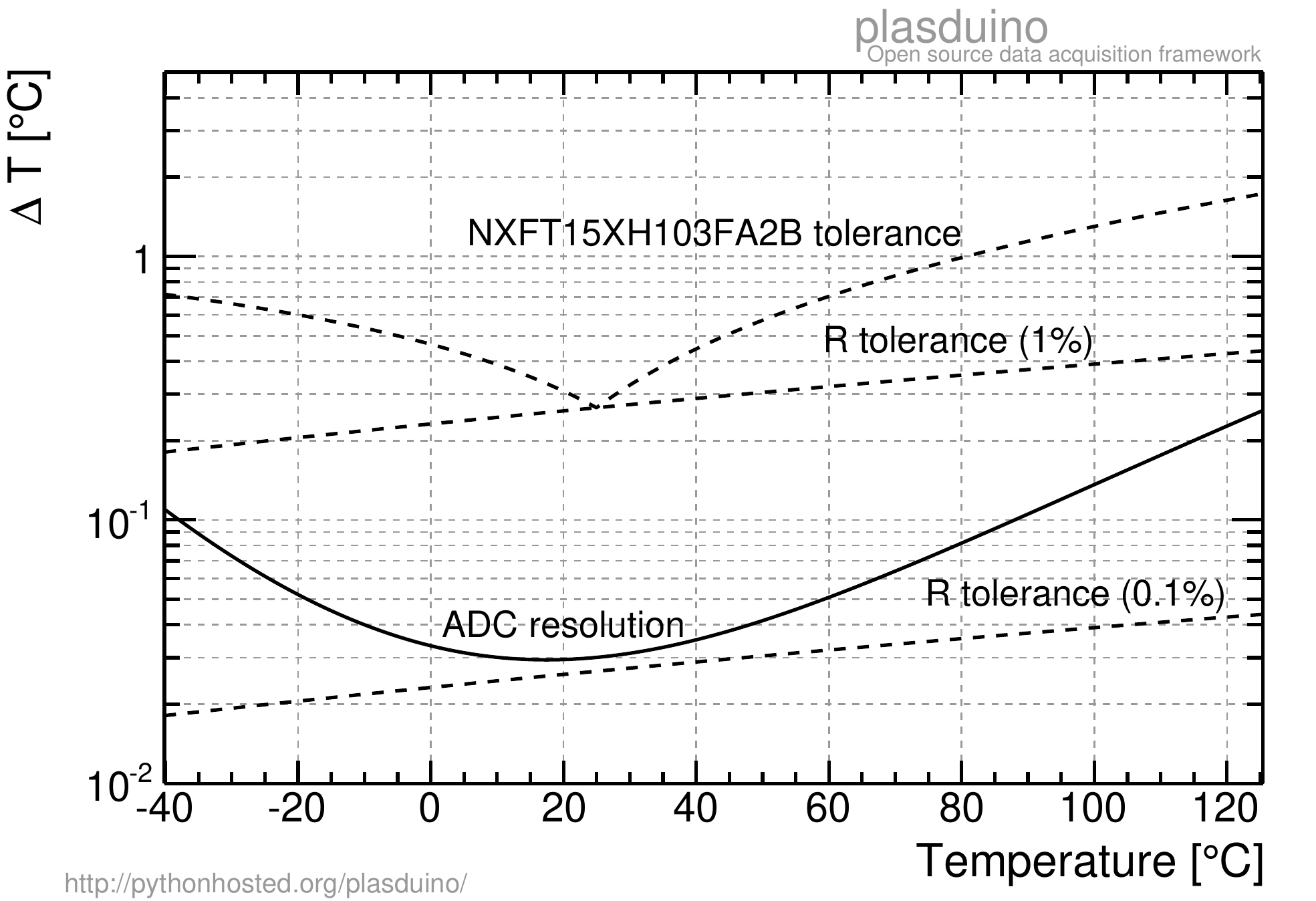}
\end{figure}

The thermistor is used in conjunction with a $10$~k$\Omega$ resistor to build a
voltage divider whose output voltage is fed into one of the \arduino\ analog
inputs. Given the $R$-$T$ characteristic of the sensor, a 10-bit ADC provides a
granularity of $\sim 0.1~^\circ$C (corresponding to a resolution of
$\sim 0.03~^\circ$C) between $0~^\circ$C and $40~^\circ$C (see
figure~\ref{fig:tempmon_resolution}).
As we shall see in section~\ref{subsec:bar} this type of differential
resolution is indeed achievable in real-life applications over reasonably
narrow temperature ranges.

The systematic uncertainty in the absolute temperature scale is dictated
by several different aspects, among which are the linearity of the ADC, the
tolerance of the auxiliary resistor (a $0.1\%$ component translates into
a typical uncertainty smaller than $0.03~^\circ$C, constituting a good compromise
between price and performance) and the deviation of the sensor response from the
nominal characteristic (typically of the order of $0.5~^\circ$C or less).
Figure~\ref{fig:tempmon_resolution} shows that an overall absolute accuracy
below $1~^\circ$C can be easily achieved with our setup.

\subsection{Rate capability}

For most of our applications we transfer each single \emph{event} (e.g., a
timestamp or an analog readout) individually to the host computer as soon as
the measurement is acquired, as this approach provides the maximal flexibility
downstream in terms of monitoring the progress of the data acquisition.
In this event-based operation mode the available bandwidth for the serial
communication is typically the limiting factor.

Assuming, for the sake of discussion, an event size of $8$--$10$ bytes
(e.g., $1$ byte for a control header, $1$ byte for the digital/analog pin
number, $4$ bytes for the timestamp, and, optionally, $2$ bytes for an ADC
reading, in addition to a $2$~bits per byte overhead due to the start and stop
bits of the serial protocol) one saturates the maximum $115200$ baud rate of the
\arduino\ interface at about $1.5$--$2$~kHz. Indeed, the system has proven
to run stable with a $\sim 2$~kHz square wave from an external pulse
generator fed into a digital input. For applications involving multiple analog
inputs one should keep in mind that they are multiplexed on a single ADC
so care must be taken to ensure that the signal at the ADC input has enough
time to settle.

We note, in passing, that if one is interested in operating the system in
counting mode (i.e., accumulating the number of counts into an internal
register for a fixed amount of time and transferring the value synchronously),
it is easily possible to achieve rates in excess of $10$~kHz.

\section{A few real-life applications}

In order to illustrate the potential of the system, in this section we describe
in detail a couple of specific educational experiments.

\subsection{The ``digital'' pendulum}

This experiment involves the use of an optical gate to study the period of a
pendulum as a function of time and/or oscillation amplitude.
The discriminated output from the optical gate is fed into an \arduino\ digital
input where each signal edge triggers an interrupt sending the timestamp to the
host PC. At the end of the acquisition session the data are post-processed
into an ASCII file for further analysis.

\begin{figure}[!htb]
  \includegraphics[width=0.5\textwidth,clip=true,trim=40 75 40 0]%
                  {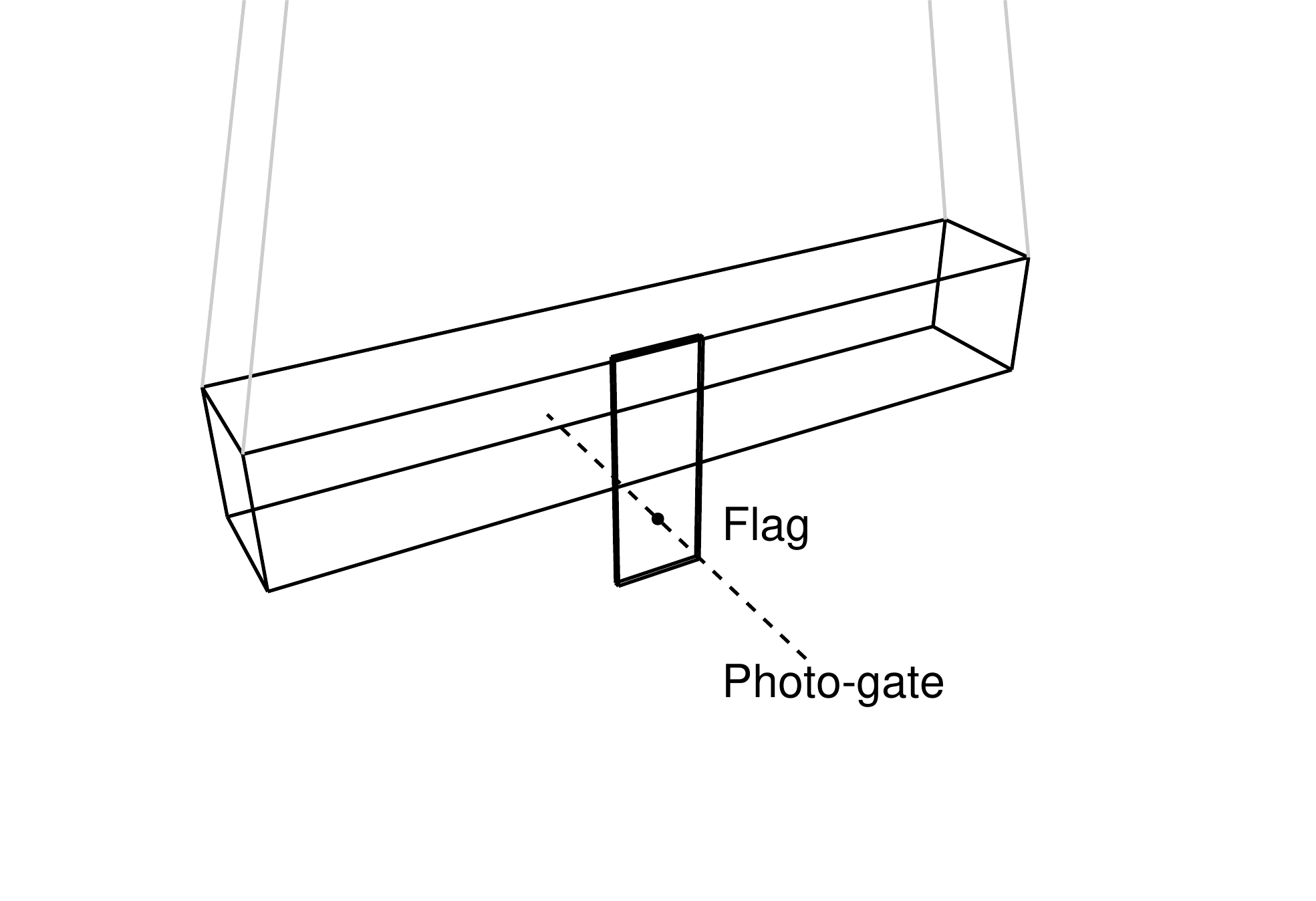}\hfill%
  \includegraphics[width=0.5\textwidth]{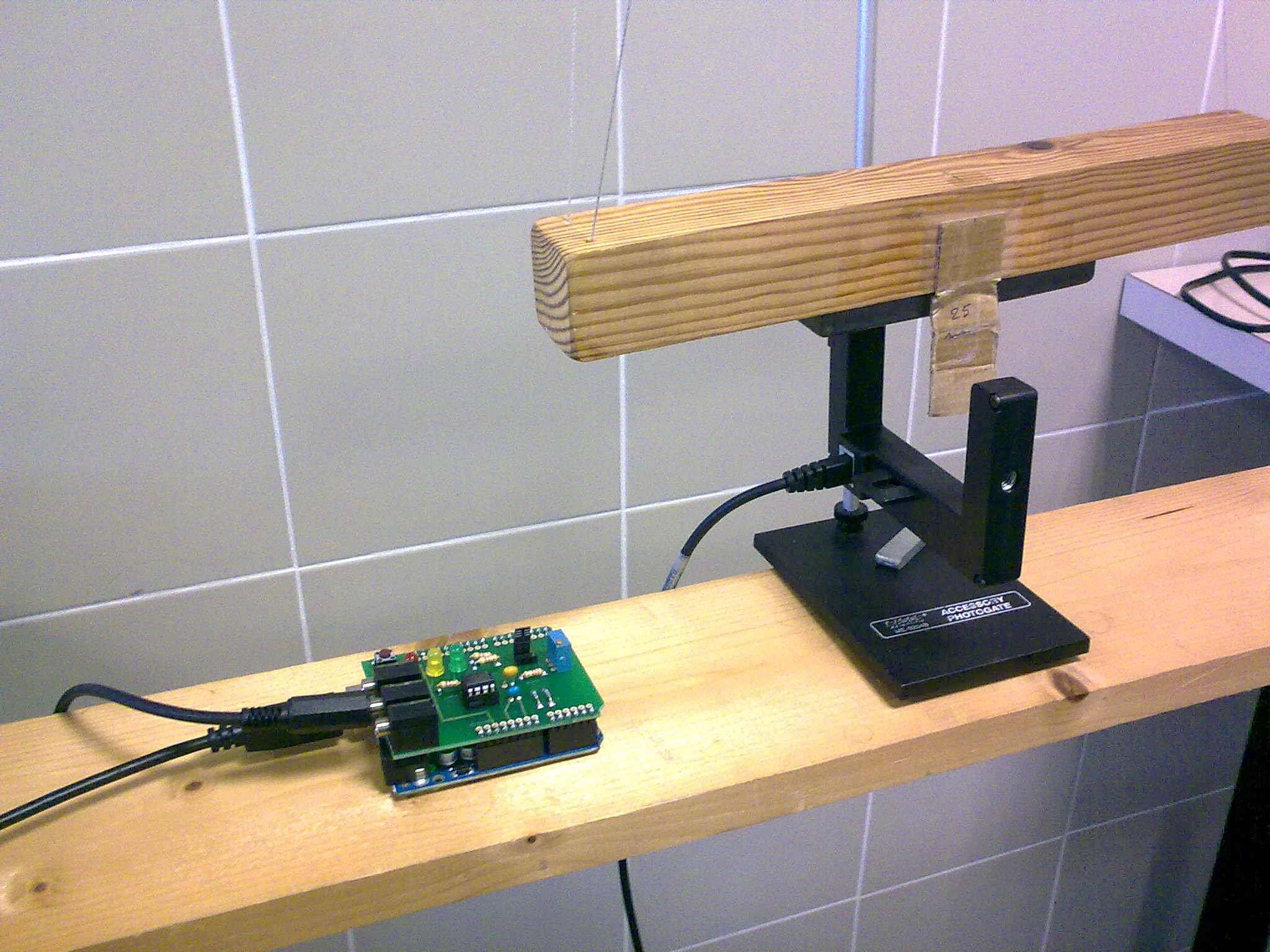}
  \caption{Schematic view (of the left) and picture (on the right)
      of our \emph{digital pendulum} setup.
      The \plasduino\ hardware (the \arduino\ board with an interface shield) 
      is also shown in the picture.
      In order to stabilize the oscillation plane, the pendulum is suspended
      with four strings. A small flag attached to its body and passing through
      an optical gate allows to measure the period and the velocity at the
      bottom of the oscillation.}
    \label{fig:pendulum_digital}
\end{figure}

Figure~\ref{fig:pendulum_digital} shows a schematic representation of the
basic setup. In order to stabilize the oscillation plane, the pendulum is 
supported by four strings. The measurement of the transit time of a small flag
attached to its body allows to recover the velocity of the center of mass at
the bottom of the oscillation---and, therefore, the total mechanical energy.
Since the $Q$ factor of the system is reasonably large
(typically $500$--$1000$), to first order one can estimate
the oscillation amplitude by imposing the conservation of energy:
\begin{equation}\label{eq:pendulum_theta}
  \theta = 2\arcsin\left( \frac{k}{T_t} \right) \quad {\rm with} \quad
  k = \left( \frac{w}{2d}\sqrt{\frac{l}{g}} \right),
\end{equation}
where $w$ is the width of the flag, $T_t$ the measured transit time, $l$ the
length of the pendulum (i.e., the distance between the suspension axis and the
center of mass) and $d$ the distance between the suspension axis and the light
beam from the photogate.

\begin{figure}[!htb]
  \includegraphics[width=0.5\textwidth]{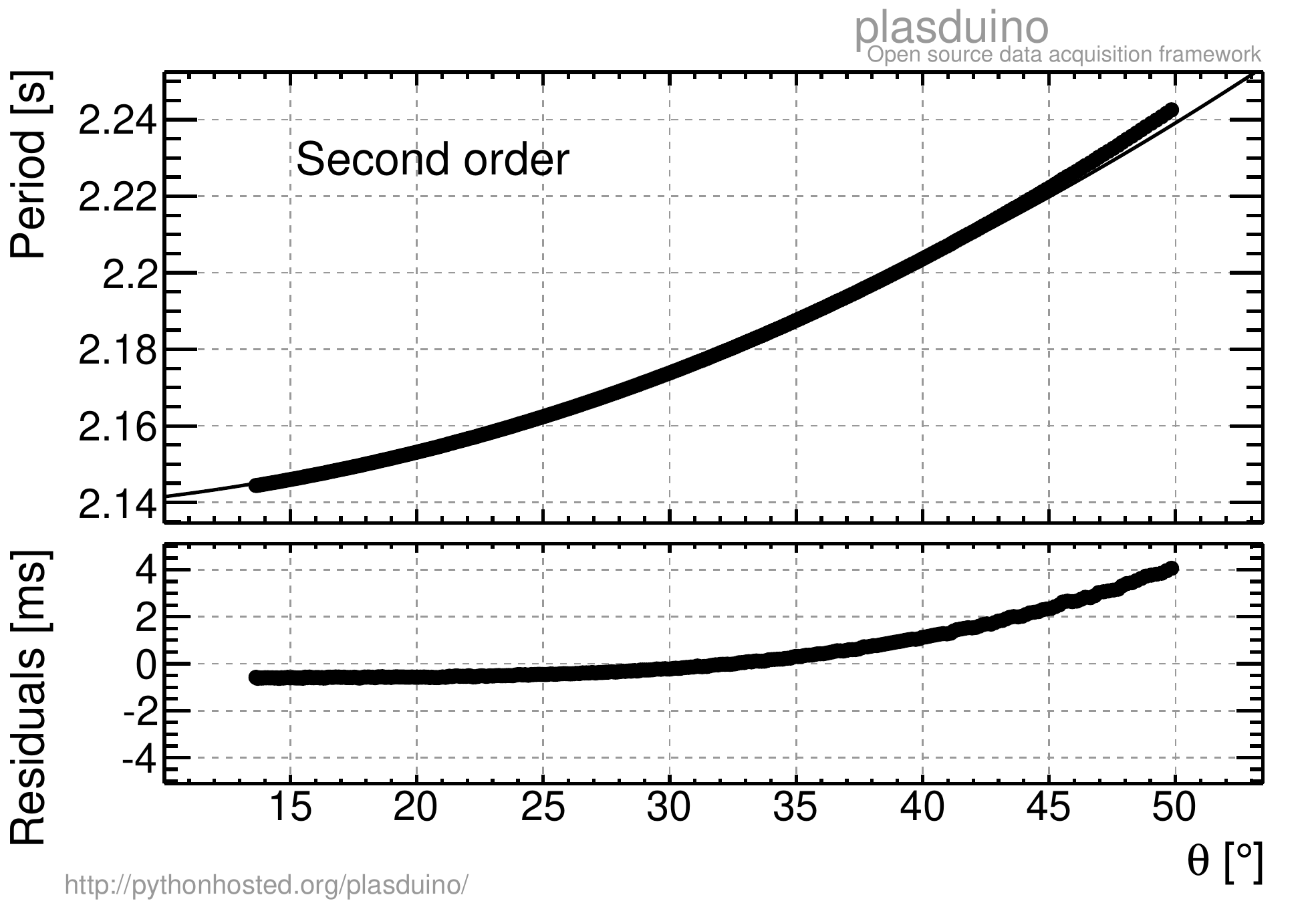}%
  \includegraphics[width=0.5\textwidth]{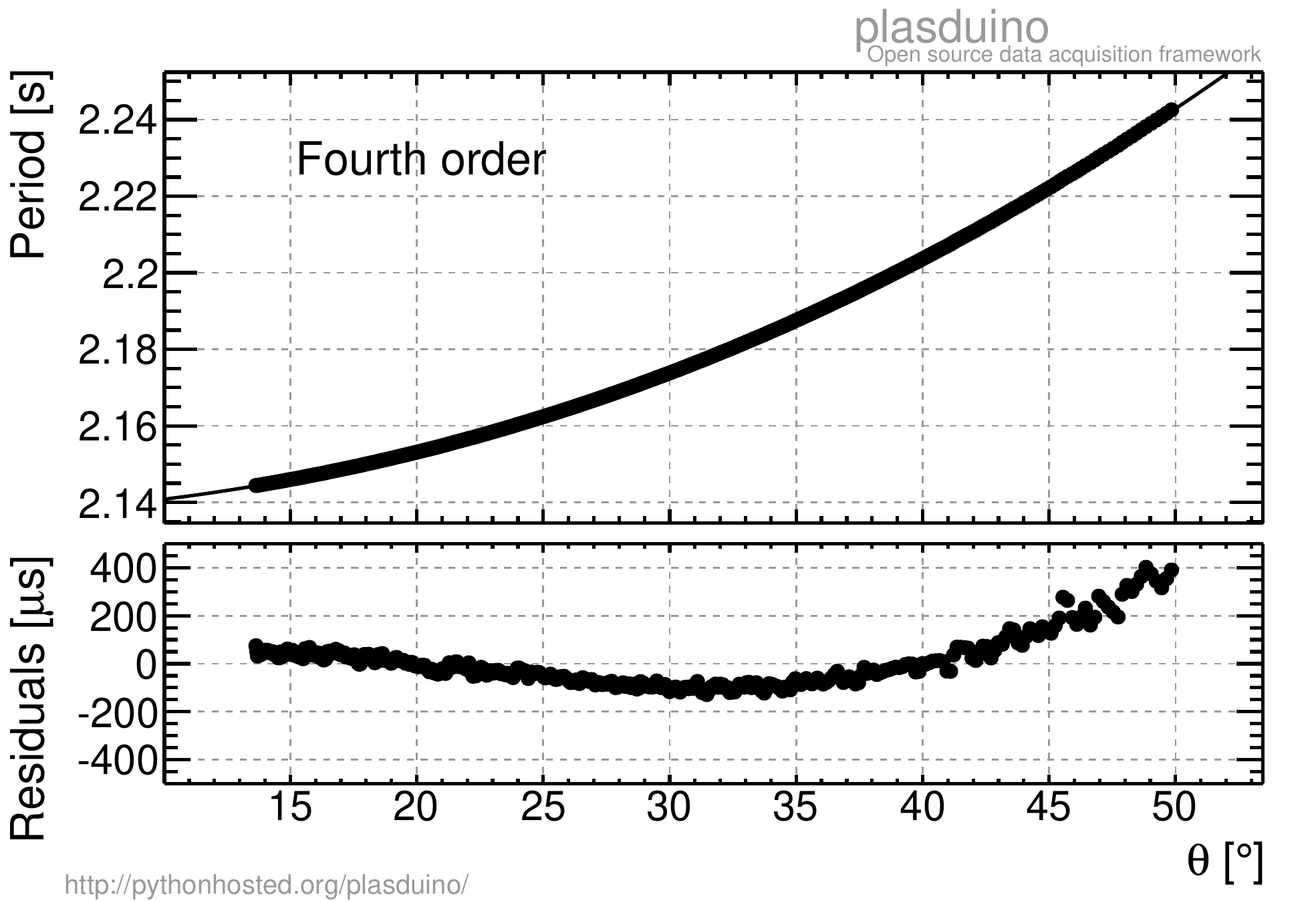}\\
  \includegraphics[width=0.5\textwidth]{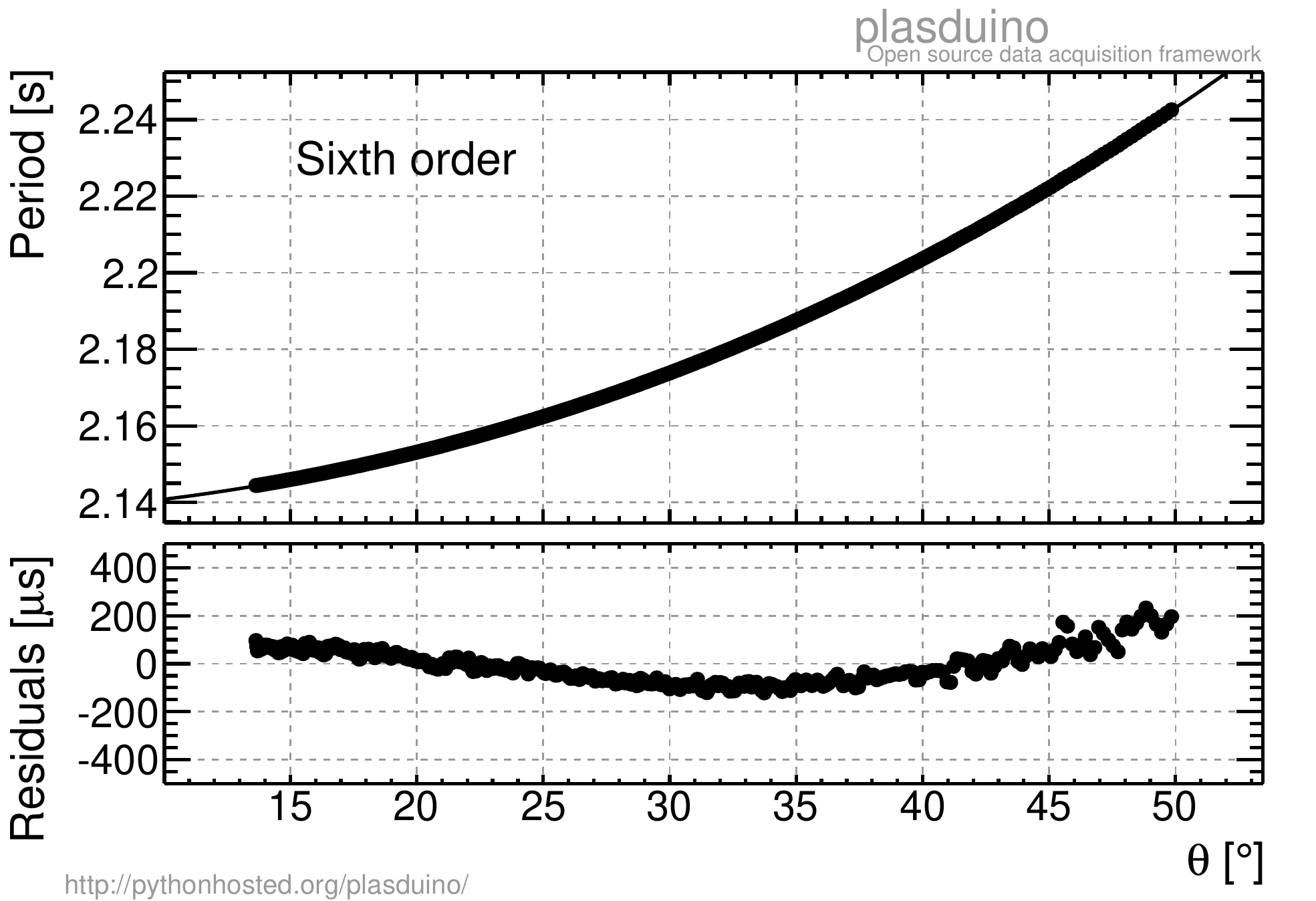}\hfill%
  \begin{minipage}[b]{0.45\textwidth}%
    \caption{Anharmonicity study with our \emph{digital pendulum} setup.
      The measured period is plotted as a function of the oscillation
      amplitude---estimated from eq.~(\ref{eq:pendulum_theta})---and fitted
      with the Taylor expansion in eq.~(\ref{eq:pendulum_period}), truncated
      at second, fourth and sixth order. In each of the three case only
      the multiplicative constant $T_0$ is left free to float in the fit.
      Note the different y-axis scales for the residual plots.}
    \label{fig:pendulum_period_amplitude}
  \end{minipage}
\end{figure}

Figure~\ref{fig:pendulum_period_amplitude} shows an illustrative example of
anharmonicity study. The measured period is plotted against the oscillation
amplitude estimated from eq.~(\ref{eq:pendulum_theta}) and fitted with the
Taylor expansion~\cite{ref:pendulum} of the period itself:
\begin{equation}\label{eq:pendulum_period}
  T = T_0 \left( 1 + \frac{1}{16}\theta^2  + \frac{11}{3072} \theta^4 +
  \frac{173}{737280} \theta^6 + \frac{22931}{1321205760} \theta^8 +
  \cdots \right),
\end{equation}
truncated at different orders (note that in every case only $T_0$ is left free
in the fit, while all the coefficients of $\theta^n$ are fixed to their nominal
values). For reference, at $50^\circ$ the relative weight of the $\theta^4$
term, from eq.~(\ref{eq:pendulum_period}), is $\sim 2.1 \times 10^{-4}$
(or $\sim 4$~ms for a $\sim 2$~s period) while that of the $\theta^6$ term is
$\sim 1.0 \times 10^{-4}$ (or $\sim 200~\mu$s for a $\sim 2$~s period).
These figures are in reasonable agreement with the results shown in
figure~\ref{fig:pendulum_period_amplitude}, though it is clear that the model
in eq.~(\ref{eq:pendulum_theta}) is too rough to reproduce the richness of the
data, as systematic residuals at the level of $\sim 100~\mu$s are
present the fit with the eq.~(\ref{fig:pendulum_period_amplitude}) truncated
at the sixth order.

\subsection{The ``analog'' pendulum}\label{subsec:analog_pendulum}

\begin{figure}[!htb]

  \begin{minipage}[b]{0.45\textwidth}%
  \includegraphics[width=\textwidth,clip=true,trim=100 40 100 55]%
                  {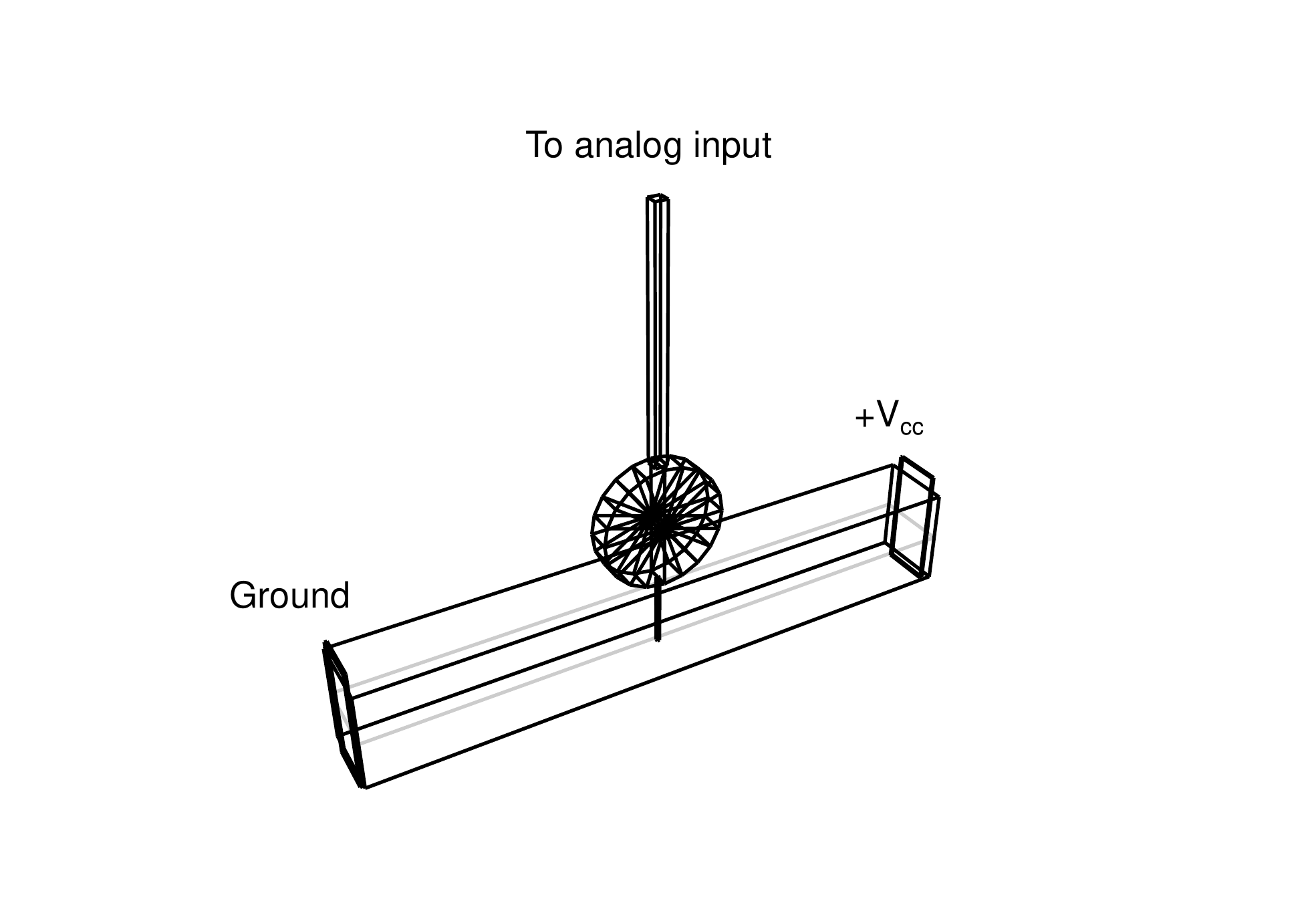}\\
    \caption{Schematic view (above) and picture (on the right)
      of our \emph{analog pendulum} setup. The water
      container and the pendulum tip, together, act as a voltage divider
      allowing to track in real time the position of the tip itself. For
      reference, the dimensions of the container are
      $43.0 \times 5.0 \times 4.5$~cm$^3$.}
    \label{fig:pendulum_analog}
  \end{minipage}\hfill%
    \includegraphics[width=0.5\textwidth]{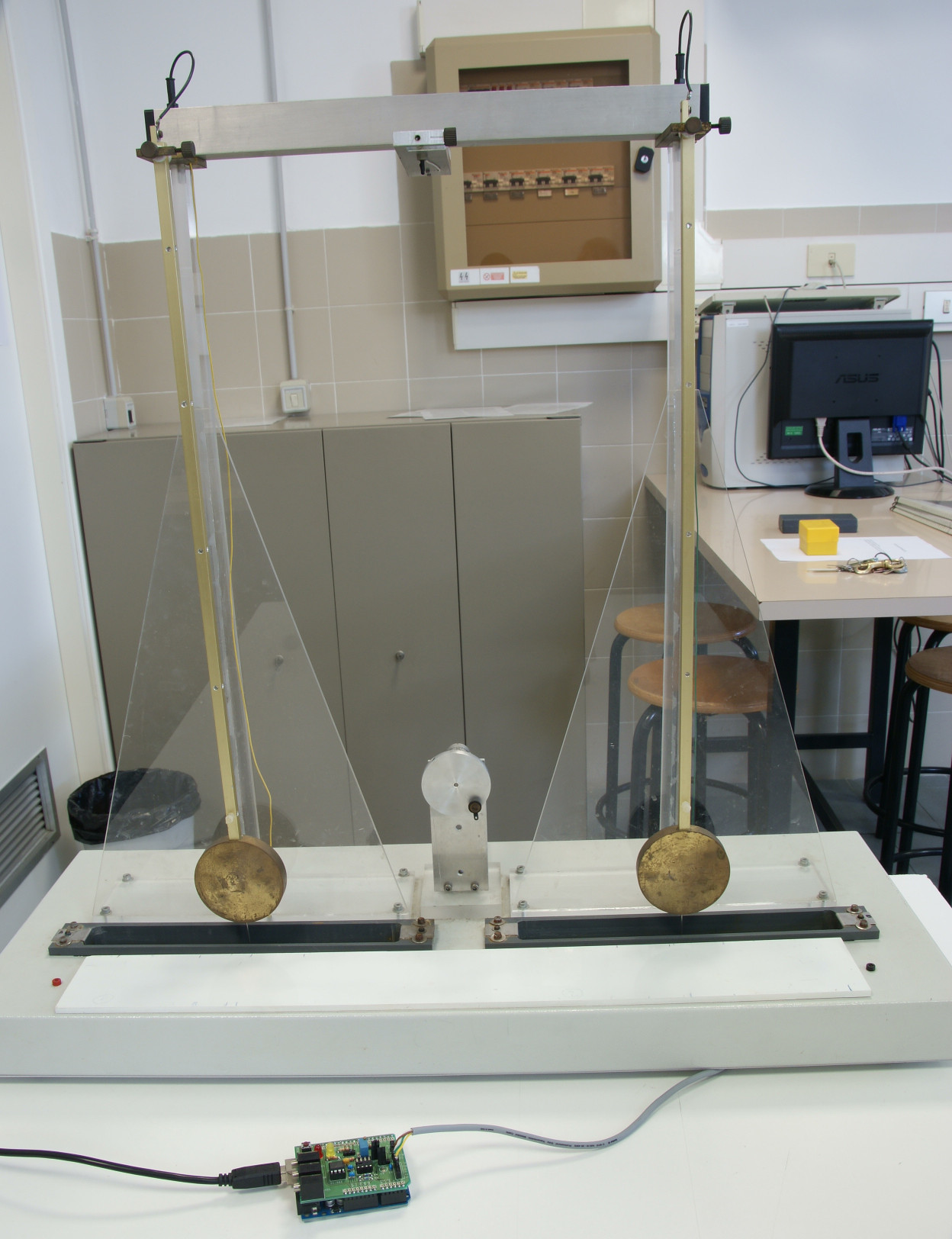}
\end{figure}

This is a variation on the theme where we use a small plastic container filled
with demineralized water as a voltage divider for a real-time tracking of the
motion of a pendulum. As shown in figure~\ref{fig:pendulum_analog}, the
container features two metal plates (one on each of the two ends) connected
to the ground and $V_{\rm cc}$ voltage references provided by the \arduino\
board. The tip of the pendulum is immersed in the water and connected to one
of the analog inputs of the board itself so that---non linearity and
geometrical projective effects aside---the voltage at the pin is proportional
to the position of the pendulum.

With a $10$~bit ADC and a maximum excursion of $\sim 40$~cm a theoretical
granularity of $40/2^{10}$~cm (or $400~\mu$m) can be achieved---and a
corresponding spatial resolution of $400/\sqrt{12} \sim 115~\mu$m.
It is quite remarkable, in fact, how such a simple setup allows to easily
achieve a sub-mm position resolution.

\begin{figure}[!htb]
  \includegraphics[width=0.5\textwidth]{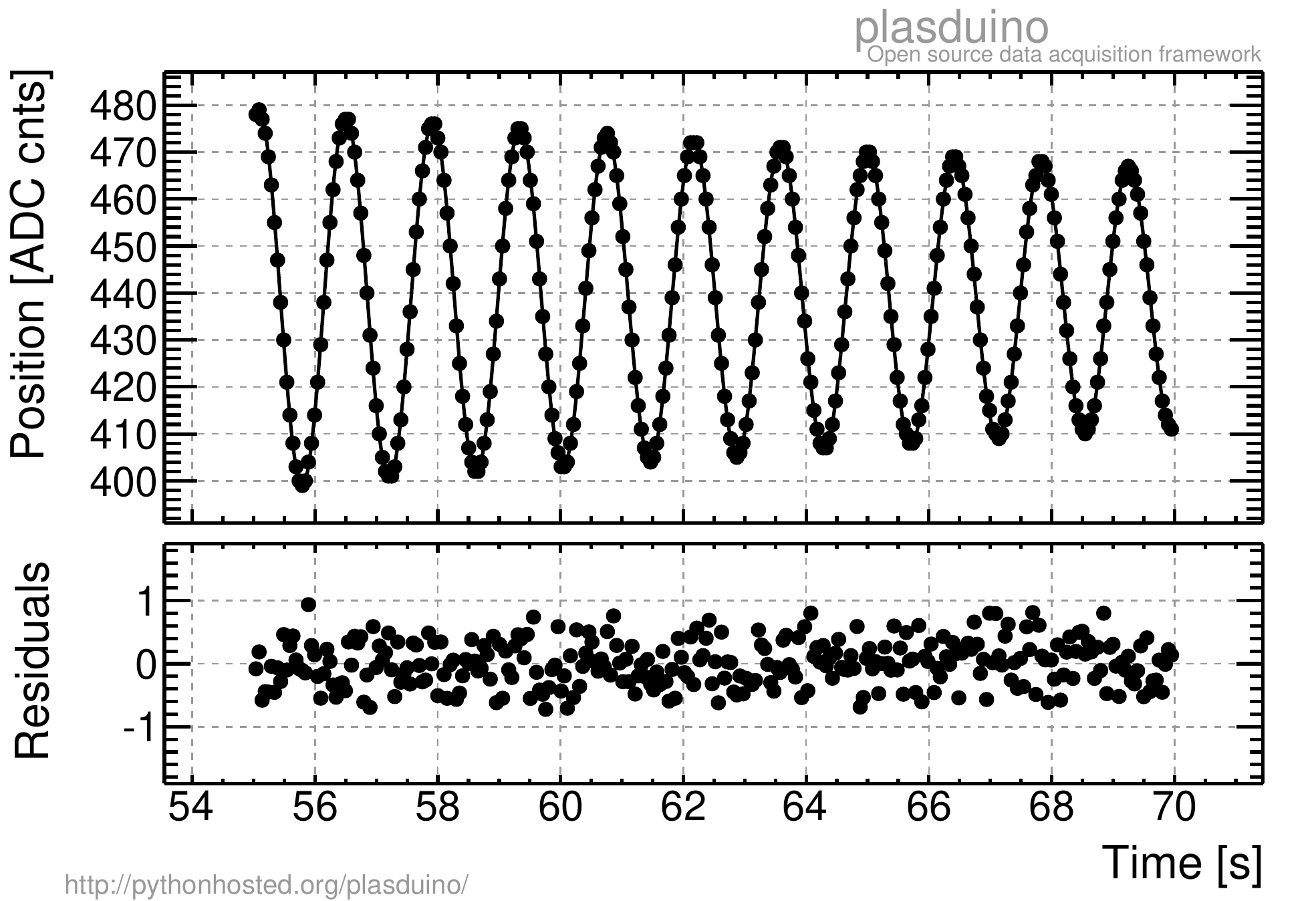}\hfill%
  \begin{minipage}[b]{0.45\textwidth}%
    \caption{Typical example of real-time pendulum tracking. A fit with
    eq.~(\ref{eq:pedulum_dump}) assuming the nominal $1/\sqrt{12}$~ADC counts
    as $1$-$\sigma$ errors on the position measurements yields a $\chi^2$ of
    $457.4/293$, indicating that the position resolution (at least in the
    $\sim 3.5$~cm range shown here) is only slightly ($\sim 25\%$) worse than
    the theoretical limit. Note that the errors on the data points and on the
    residuals are omitted for visualization purposes.}
    \label{fig:pendulum_dump}
  \end{minipage}
\end{figure}

Figure~\ref{fig:pendulum_dump} shows a sample data acquisition, with the
experimental points fitted with the function
\begin{equation}\label{eq:pedulum_dump}
  x(t) = c_0 + c_1 e^{-\lambda t}\sin(\omega_0 t + \phi_0).
\end{equation}
The root mean square of the residuals is $\sim 0.36$~ADC counts, which is only
$25\%$ larger that the nominal $1/\sqrt{12}$ value one would obtain if
the system noise and non-linearity were negligible. (We stress, however,
that non linearities at the level of a few \%, that can possibly be handled
with a dedicated calibration, are observed at the edges of the container.)

The setup illustrated in figure~\ref{fig:pendulum_analog} can be easily
modified as to include a small motorized aluminum wheel with a fixed pin
connected to the body of the pendulum through a pair of springs and acting as
a(n approximately) sinusoidal external force. The speed of the motor (and hence
the frequency of the input load) can be controlled by means of the \arduino\
PWM capabilities (and, assuming that the apparatus is properly sized, the
current provided by a standard USB port is enough to drive the motor and no
external power supply is required).

\begin{figure}[!htb]
  \includegraphics[width=0.5\textwidth]{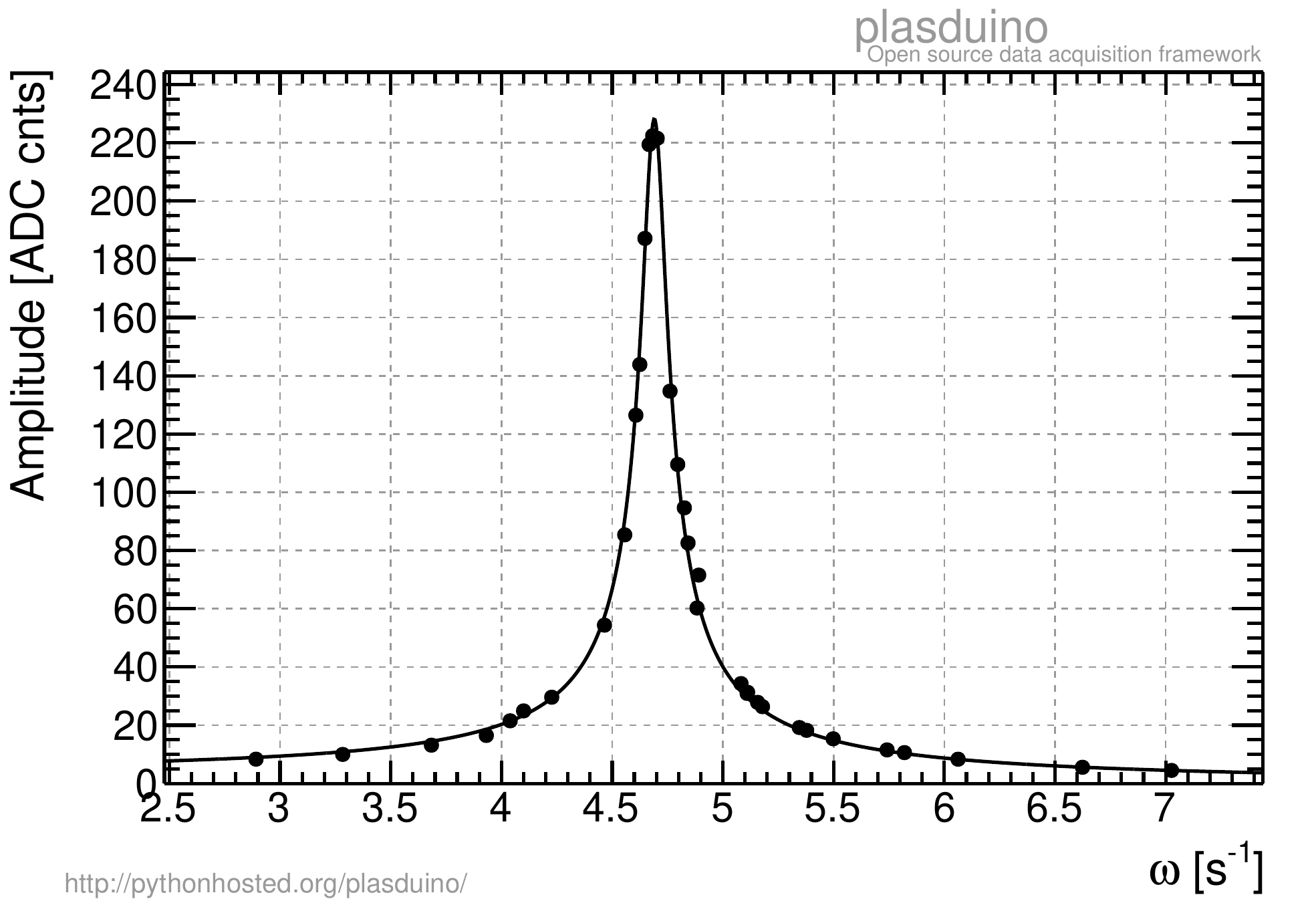}\hfill%
  \begin{minipage}[b]{0.45\textwidth}%
    \caption{Resonance curve of a pendulum driven by an external
    sinusoidal force, produced using a slightly modified version of the
    setup illustrated in figure~\ref{fig:pendulum_analog}.
    The data points are fitted with the function in eq.~(\ref{eq:resonance}).
    We note that no attempt has been made to calibrate the non linearity
    of the system.}
    \label{fig:pendulum_resonance}
  \end{minipage}
\end{figure}

This modified setup allows to study the steady-state oscillation amplitude of
the pendulum as a function of the angular pulsation of the external force,
producing a resonance curve of the system.
Figure~\ref{fig:pendulum_resonance} shows how, even with no attempt at
calibrating the non linearities, the system response is in remarkable
agreement with the prediction of the simplest possible model (i.e., with a
single friction term proportional to the velocity): 
\begin{equation}\label{eq:resonance}
  A(\omega) = \frac{A_0}{(\omega_0^2 - \omega^2)^2 + 4\Gamma^2\omega^2}.
\end{equation}

\subsection{Measurement of thermal conductivity}\label{subsec:bar}

\begin{figure}[!htb]
  \includegraphics[width=0.48\textwidth]{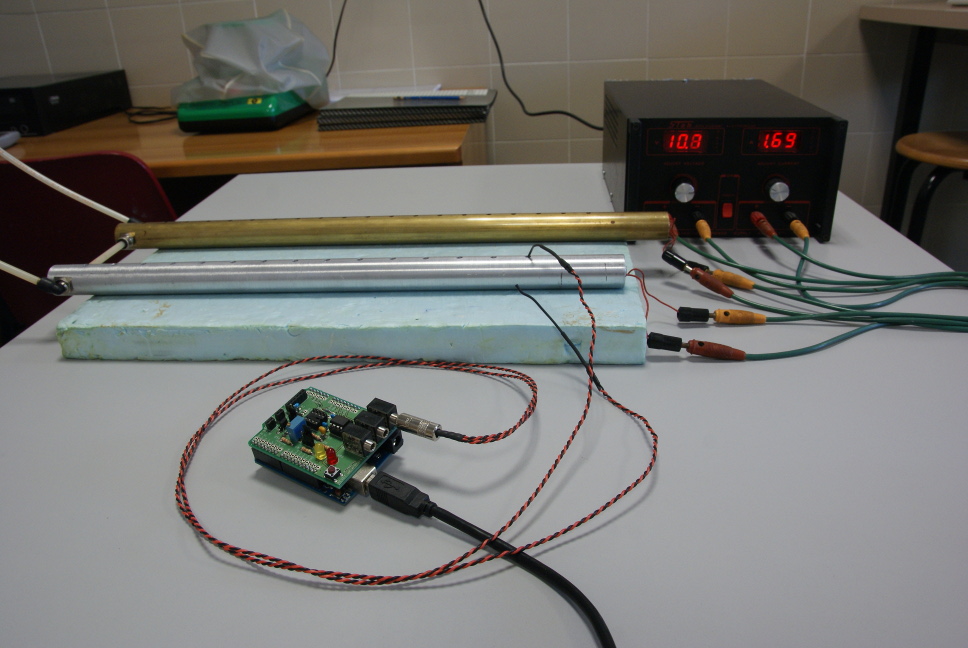}\hfill%
  \includegraphics[width=0.5\textwidth]{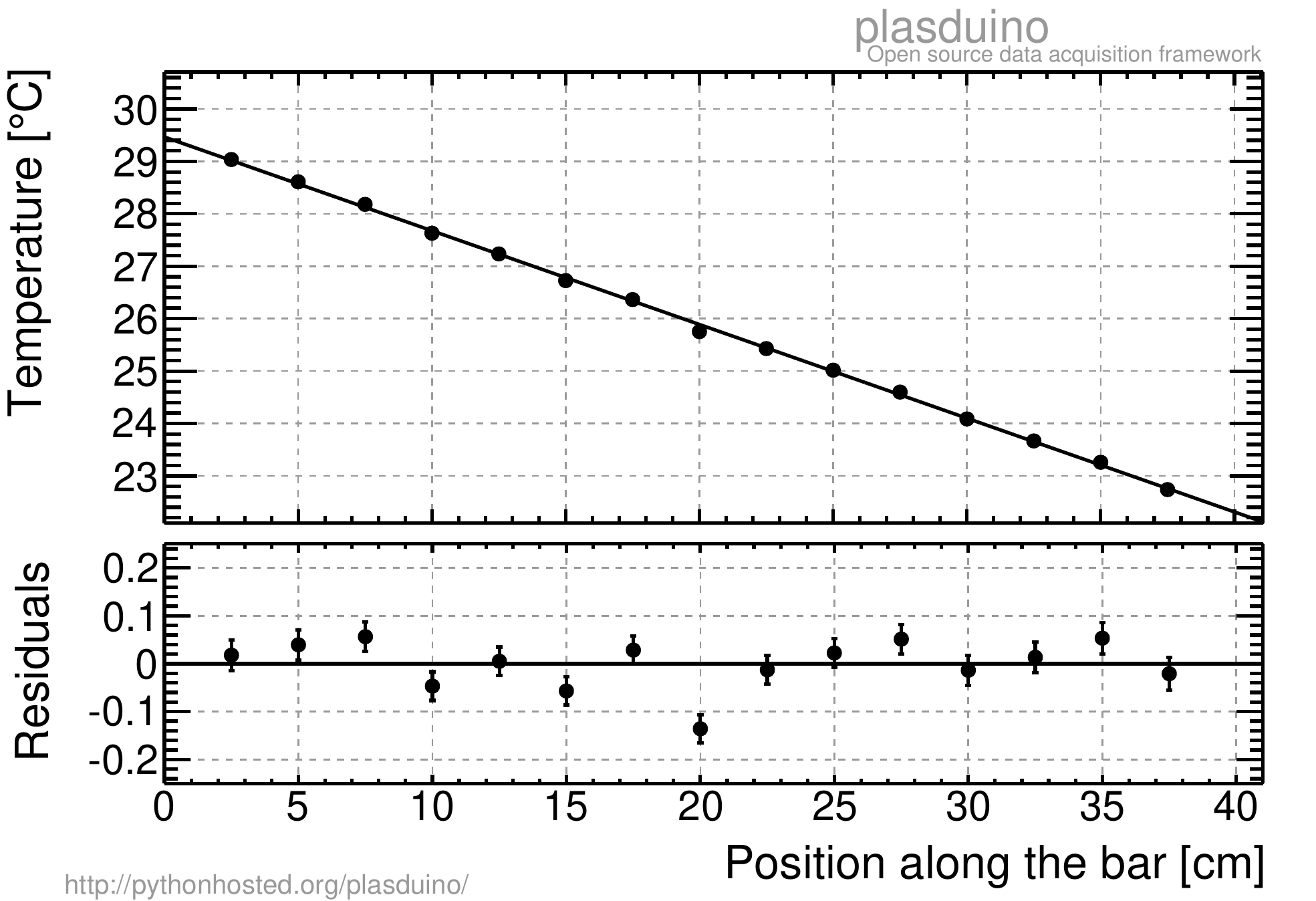}
  \caption{On the left, the picture of the apparatus showing also the plasduino
    system connected with the two thermistors.
    On the right, the temperature profile of a metal bar kept 
    at constant (different) temperatures at its extremes. 
    A linear fit assuming that the measurement
      errors are solely dictated by the ADC resolution (i.e.,
      $\sim 0.03~^\circ$C) yields a $\chi^2$ of $35.9/13$, possibly indicating
      that the errors themselves are slightly underestimated. For reference,
      the relative uncertainty on the fitted slope is $\sim 0.5\%$.}
    \label{fig:tempmon}
\end{figure}

We illustrate this thermodynamics experiment as a last illustrative example of
the capabilities of the system. The setup is as simple as a metal bar with a
regular set of holes allowing to measure the temperature profile as a function
of the position along the bar itself. The temperatures on the two ends of the
bar are set by a resistor heated up due to Joule effect and a steady flow of
running fresh water, respectively. 
The left panel of figure~\ref{fig:tempmon} shows the apparatus, with 
two bars of different material available. The plasduino board is also visible 
with two thermistors connected. One thermistor is placed in the first hole of 
one bar, while the other is left floating to measure air temperature.
Neglecting the heat exchanges with the
ambient (which can be minimized by ensuring that the temperature of any point
in the bar is not \emph{too} different from the ambient temperature), 
the theory predicts the temperature profile to be linear
\begin{equation}
  T(x) = T_0 - \frac{W}{\lambda S} x
\end{equation}
(here $W$ is the power provided by the voltage supply driving the resistor,
$S$ is the cross section of the bar and $\lambda$ the thermal conductivity of
the material). 
Moving the thermistor from one hole to the other, the temperature profile
can be measured, as shown in the right panel of figure~\ref{fig:tempmon}, 
and its slope allows to estimate $\lambda$.

\subsection{Displaying data in real time}

\begin{figure}[!htb]
  \centering\includegraphics[width=\textwidth]{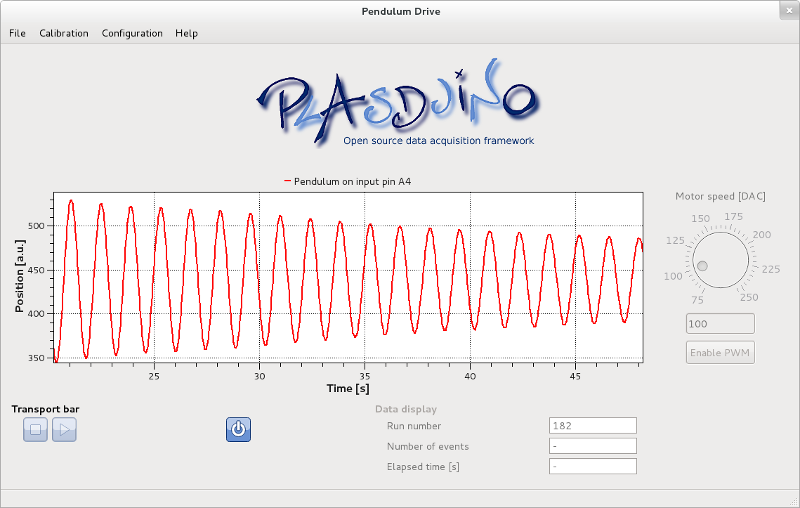}

  \caption{Graphical user interface for the analog pendulum setup described
  in section~\ref{subsec:analog_pendulum}. The canvas displaying the time
  series can be used interactively and allows to perform the basic analysis
  previously described with the computer mouse---without any post-processing of
   the data.}
  \label{fig:pendulumdrive}
\end{figure}

Admittedly, \plasduino\ is designed as a \emph{data acquisition system}---i.e.
emphasis is put on the two basic tasks of (i) collecting data and (ii) storing
them in a suitable form for off-line analysis. That being said, we fully
realize the educational value of being able to display data in real time and
provide a set of interactive graphical widgets for the purpose, which are used
in some of the modules, as shown in figure~\ref{fig:pendulumdrive}.
There is no doubt this is one of the aspects where the system could be
improved, partially evolving toward an on-line learning platform.

\section{Status and prospects}

The \plasduino\ framework has been in use in the didactic laboratories of the
Physics Department of the University of Pisa for more than two years, now, in
the form of a set of seven educational experiments that we routinely propose to
undergraduate and high-school students---i.e., students with no data acquisition
experience and facing basic data analysis for the first time.
The graphical user interface, and the system in general, proved to be intuitive
enough to be effectively used by novices with little or no need for very
specific instructions. As a matter of fact, we have produced a number of
fully-assembled systems which are currently in use in the didactic laboratories
of a few local high schools and we have agreements with other schools to
continue working in this direction.

We strive for making available in a terse form all the information necessary to
replicate the system (electronics schematics, part lists, cable and connector
layouts, software installation instructions) and we sustain that this can be
realistically achieved by anybody with basic computer skills and some
soldering experience (e.g., many of the high-school technicians we interacted
with during the development).
It goes without saying that actively developing the system in any of its parts
requires skills that go well beyond those mentioned in the previous paragraph.

In the near term we are working on the implementation of a few more advanced
educational modules, most notably a configurable waveform generator and a
data acquisition shield to read signals from photomultiplier tubes
(integrating a GPS receiver for the absolute timestamp). In parallel, we started
a collaboration with a local high school to assemble a fully fledged weather
station.

\acknowledgments

We gratefully acknowledge the continuous support to this project of the Physics
Department ``E.~Fermi'' of the University of Pisa.

We also are grateful to Philippe Bruel, Ric Claus, Scilla Degl'Innocenti
and Francesco Longo for their encouragement in the initial phase of the
project.

\medskip

\noindent The \plasduino\ team is listed at
\url{http://pythonhosted.org/plasduino/team.html}.

\end{document}